\PassOptionsToPackage{svgnames}{xcolor}
\documentclass{vgtc}

\graphicspath{{figures/}{pictures/}{images/}{./}}

\usepackage{times}

\usepackage{amssymb}
\usepackage{tabu}
\usepackage{booktabs}
\usepackage{amsmath}
\usepackage{multirow}
\usepackage{csquotes}
\usepackage{graphicx}
\usepackage{epstopdf}
\usepackage{svg}
\usepackage{pgfplots}
\pgfplotsset{compat=1.18}
\usepackage{pgfplotstable}
\usepackage{booktabs}
\usepackage{siunitx}
\onlineid{0}
\vgtccategory{Research}
\vgtcinsertpkg

\title{Multimodal Feedback for Task Guidance in Augmented Reality}

\author{%
  Hu Guo\thanks{first author, e‑mail: goudn@mail.uc.edu}  \scriptsize Department of Computer Science, University of Cincinati, USA
  \and Lily Patel\scriptsize Department of Human–Computer Interaction, University of Cincinati, USA
  \and Rohan Gupta\thanks{corresponding author, e‑mail: rohan.gupta@mail.uc.edu}\scriptsize Department of Mechanical Engineering, University of Cincinati, USA
}

\abstract{%
Optical see‑through augmented reality (OST‑AR) overlays digital targets and annotations on the physical world, offering promising guidance for hands‑on tasks such as medical needle insertion or assembly. Recent work on OST‑AR depth perception shows that target opacity and tool visualization significantly affect accuracy and usability; opaque targets and rendering the real instrument reduce depth errors whereas transparent targets and absent tools impair performance. However, reliance on visual overlays may overload attention and leaves little room for depth cues when occlusion or lighting hampers perception. To address these limitations, we explore multimodal feedback that combines OST‑AR with wrist‑based vibrotactile haptics.

The past two years have seen rapid advances in haptic technology. Researchers have investigated skin‑stretch and vibrotactile cues for conveying spatial information to blind users, wearable ring actuators that support precise pinching in AR, cross‑modal audio‑haptic cursors that enable eyes‑free object selection, and wrist‑worn feedback for teleoperated surgery that improves force awareness at the cost of longer task times. Studies comparing pull versus push vibrotactile metaphors found that pull cues yield faster gesture completion and lower cognitive load. These findings motivate revisiting OST‑AR guidance with a fresh perspective on wrist‑based haptics.

We design a custom wristband with six vibromotors delivering directional and state cues, integrate it with a handheld tool and OST‑AR, and assess its impact on cue recognition and depth guidance. Through a formative study and two experiments (N=21 and N=27), we show that participants accurately identify haptic patterns under cognitive load and that multimodal feedback improves spatial precision and usability compared with visual‑only or haptic‑only conditions. Participants report greater confidence and reduced cognitive effort, albeit with slightly longer completion times. Our findings suggest that wrist‑based haptics can supplement OST‑AR to mitigate depth perception challenges, reduce visual workload and support high‑precision manual tasks. We discuss design guidelines informed by recent haptic research and highlight implications for education, simulation and cross‑modal interfaces.
}

\keywords{Augmented reality, haptics}

\begin{document}
\firstsection{Introduction}
\maketitle

\subsection{Background and Motivation}
Optical see-through augmented reality (OST-AR) enables users to superimpose virtual objects onto the real world, providing intuitive guidance for tasks ranging from surgical procedures to manufacturing assembly. In such systems, accurate depth perception is crucial since misjudging a virtual target’s position can lead to misalignment or tissue damage. Recent work by Yang et al. systematically evaluated how target transparency and tool visualization influence depth perception and usability in OST-AR \cite{Yang2020DepthPerception}. They found that opaque targets yielded smaller depth errors than transparent ones and that visualizing the real surgical instrument provided the highest localization accuracy and usability \cite{Yang2020DepthPerception}. Conversely, when the tool was not visualized, participants overshot the transparent target and reported increased workload \cite{Yang2020DepthPerception}. 

Despite advances in display technology, reliance on visual overlays alone can saturate the user’s attentional resources due to complex scenes, lighting variations or occlusion by the user’s hands and instruments. Prior research in haptic guidance suggests that tactile feedback can reduce visual workload by offloading directional cues to the somatosensory system \cite{Weber2008VibrotactileGuidance}. In mock surgical tool experiments, vibrotactile patterns on a handheld handle enabled sub-millimeter accuracy after training but were cognitively demanding due to ambiguous vibrations \cite{Weber2008VibrotactileGuidance}. Wrist-worn devices offer a compelling alternative: they leave the hands free for tool manipulation and can convey spatial information through the natural mapping between skin regions and directions \cite{Oyarzun2019VibrotactilePuzzle}. Wristbands and bracelets have been used to provide vibrotactile cues for 2D puzzles \cite{Oyarzun2019VibrotactilePuzzle}, notify users of successful hand gestures and compensate for AR tracking errors \cite{Oyarzun2019VibrotactilePuzzle}, and improve the realism of virtual interactions by combining squeeze and vibration \cite{Pezent2021Tasbi}. Recent developments have further diversified haptic modalities. For example, Li et al. showed that blind VR users more accurately understood object location and movement when skin-stretch cues were applied to the dorsal hand compared with vibrotactile cues \cite{Li2025SkinStretch}, suggesting that alternative forms of haptic feedback may convey spatial information more effectively. Meanwhile, low-profile haptic rings for finger pinching produced greater precision and faster completion times than glove-based systems in an AR object manipulation task \cite{Nguyen2025PinchRing}, demonstrating that simple wearable devices can outperform more complex hardware.

\subsection{Research Questions and Contributions}
Building upon the depth perception study of Yang et al. \cite{Yang2020DepthPerception}, we explore how vibrotactile haptic feedback can supplement OST-AR for handheld tool guidance. Specifically, we address the following research questions:
\begin{itemize}
  \item \textbf{RQ1:} How accurately can users identify directional and state-based haptic cues delivered via a wrist-worn vibrotactile device while performing a visual search task? Prior work indicates that vibration patterns are intuitive when mapped to body coordinates \cite{Oyarzun2019VibrotactilePuzzle}, but distinguishing subtle differences can be challenging \cite{Weber2008VibrotactileGuidance}. Recent comparisons of pull versus push metaphors for vibrotactile guidance showed that “pull” cues yield faster gesture completion and lower cognitive load than “push” cues \cite{Smith2023PullPush,Smith2023PullPush}. We therefore adopt a pull-oriented mapping for our cues.
  \item \textbf{RQ2:} Does combining wrist-based haptics with OST-AR improve spatial precision, depth perception and subjective usability compared with using either modality alone? Haptic feedback has been shown to improve perceived stiffness and interaction realism in VR/AR \cite{Pezent2021Tasbi}, and to help users align virtual objects in 2D tasks \cite{Oyarzun2019VibrotactilePuzzle}. More recent work on teleoperated surgery demonstrates that wrist-worn haptic feedback reduces force error though it may increase task time \cite{Wang2025Teleoperation}, highlighting a speed–accuracy trade-off. We aim to quantify these effects in a handheld tool context.
  \item \textbf{RQ3:} What are users’ subjective experiences regarding cognitive load, confidence and preferences when using multimodal feedback? Previous studies report that vibrotactile cues can reduce visual workload \cite{Weber2008VibrotactileGuidance} but may be cognitively demanding to interpret \cite{Weber2008VibrotactileGuidance}. Skin-stretch feedback may further enhance spatial perception \cite{Li2025SkinStretch}, while minimalist ring-based devices achieve superior precision with lower complexity \cite{Nguyen2025PinchRing}. We examine whether our wrist-based design achieves a favorable balance.
\end{itemize}
Our contributions are threefold. First, we design a lightweight wrist-worn haptic device with six vibromotors arranged radially around the wrist, capable of conveying directional (left/right/forward/back) and state (approach/hit) cues. We adopt a pull-oriented mapping inspired by recent vibrotactile guidance research \cite{Smith2023PullPush,Smith2023PullPush}. Second, we integrate this device into an OST-AR platform for handheld tool guidance and propose a cue mapping informed by a formative study with domain experts. Third, we evaluate the system through a cue identification experiment and a tool guidance task, demonstrating that multimodal feedback improves depth accuracy and usability relative to visual-only or haptic-only conditions. We further situate our findings within the broader landscape of haptic technologies, highlighting complementary approaches such as skin-stretch feedback \cite{Li2025SkinStretch}, ring-based haptics \cite{Nguyen2025PinchRing} and audio-haptic cursors \cite{Lu2024SonoHaptics}.

\section{Foundations and Related Work}
\subsection{Depth Perception with Optical See--Through Displays}
Accurate depth perception in optical see--through AR (OST--AR) hinges on consistent occlusion, correct vergence/accommodation cues, and stable registration between virtual and physical content. Yang et~al.\ provide one of the most systematic examinations to date of these factors in a surgical guidance setting, showing that low--transparency (effectively opaque) targets combined with faithful visualization of the \emph{real} tool substantially reduce depth error and improve subjective usability \cite{Yang2020DepthPerception}. When transparency increased, participants tended to overshoot and reported higher workload, a pattern that worsened if the real tool was not visualized \cite{Yang2020DepthPerception}. The study concludes by motivating the addition of nonvisual cues and richer occlusion strategies to mitigate ambiguity in cluttered or poorly lit scenes \cite{Yang2020DepthPerception}.

In parallel, researchers have explored nonvisual channels to offload attentional demand from vision during precise alignment. Early work on vibrotactile guidance mounted on handheld tools demonstrated that tactile cues can reduce visual load and even enable sub--millimeter performance after training, albeit with notable confusion for vertically oriented cues due to tactor placement and cognitive demand \cite{Weber2008VibrotactileGuidance}. This limitation suggests that moving haptic signaling away from the grasping hand (e.g., to the wrist) may alleviate perceptual interference with fine motor control while preserving spatial expressivity.

\subsection{Vibrotactile and Multimodal Wearables}
Wrist--worn haptics are attractive because they preserve manual dexterity and map intuitively to egocentric directions around the body. Studies using vibrotactile wristbands for AR object manipulation report improved spatial awareness in 2D tasks and show that simple, peripheral cues can compensate for transient AR misalignment or gesture recognition uncertainty \cite{Oyarzun2019VibrotactilePuzzle}. Preferences also tend to favor lightweight wearables over glove--based systems when comparable guidance performance is achievable \cite{Oyarzun2019VibrotactilePuzzle}. Complementing this trend, cross--modal devices such as Tasbi combine vibrotactile actuators with controllable squeeze, improving perceived stiffness and interaction realism in VR/AR \cite{Pezent2021Tasbi}; industry reports from Meta highlight similar benefits and design trade--offs for wrist squeeze and vibration in AR/VR interactions \cite{Meta2021TasbiBlog}.

Alternative haptic mechanisms are gaining traction for spatial communication. Li et~al.\ show that skin--stretch cues applied on the dorsal hand enable more accurate perception of object position and motion than vibration for blind VR users, pointing to the dorsal surface as an informative yet unobtrusive site for directional encoding \cite{Li2025SkinStretch}. Compact proprioceptive feedback solutions further indicate that wrist rotation and hand aperture information can be conveyed with fewer actuators without degrading performance, simplifying hardware while retaining utility \cite{Choi2024Proprioception}. For fine object manipulation, ring--based actuation can outperform glove systems in precision and speed during AR pinching tasks, suggesting that low--profile, localized actuators may deliver superior control with less encumbrance \cite{Nguyen2025PinchRing}. 

Design of the \emph{coding scheme} matters as much as hardware form. A 24--actuator wristband study found that \emph{pull} metaphors (signals that seem to draw the hand toward a goal) yield faster completion times, higher usability, and lower mental demand than \emph{push} metaphors \cite{Smith2023PullPush}. In teleoperation, relocating cues to the wrist can reduce force error during robotic surgery even if task time increases, underscoring a practical speed--accuracy trade--off that designers must balance \cite{Wang2025Teleoperation}. Beyond touch alone, audio--haptic mappings like SonoHaptics align object properties (e.g., size, lightness) to sound and vibration, enabling eyes--free, lower--effort selection and hinting at complementary channels for AR guidance \cite{Lu2024SonoHaptics}. Emerging shoulder--mounted and reel--based wearables expand this palette of sensations and placements for mobile MR scenarios \cite{CMU2023ReelFeel}.

\subsection{Design Implications for AR Tool Guidance}
From this literature, three implications follow. \textbf{(1) Preserve visual veridicality but plan for failure:} opaque targets and real--tool rendering improve depth judgments \cite{Yang2020DepthPerception}, yet real deployments face occlusions, reflectance changes, and tracking dropouts. \textbf{(2) Move cues to free skin and leverage embodied mappings:} the wrist and dorsal hand offer unobtrusive surfaces with natural left/right/forward/back correspondences \cite{Oyarzun2019VibrotactilePuzzle,Li2025SkinStretch, duan2025localization}. \textbf{(3) Encode direction with pull--style, brief pulses:} pull metaphors reduce cognitive load \cite{Smith2023PullPush}; short bursts limit adaptation while reserving distinct patterns for \emph{state} confirmations \cite{Weber2008VibrotactileGuidance,Pezent2021Tasbi}. These principles inform our system architecture and cue set.

\section{System Architecture}
\subsection{Wearable Hardware}
Our multimodal guidance platform integrates an optical see--through headset (HoloLens~2), a tracked handheld stylus, and a custom six--motor wristband. The band uses a flexible silicone strap and six coin vibromotors arranged at \(60^\circ\) intervals to cover cardinal and diagonal directions. A low--mass controller (ESP32) communicates with the AR app over Bluetooth at 100~Hz and drives each channel with PWM intensity control (0--255). The assembly weighs 35~g and fits 15--20~cm wrists. Consistent with teleoperation findings that peripheral, lightweight devices minimize interference \cite{Wang2025Teleoperation} and with ring--based evidence favoring compact actuation for precision \cite{Nguyen2025PinchRing}, we prioritized low inertia and unobtrusiveness.

\subsection{Cue Policies and Feedback Mapping}
We encode \emph{directional} and \emph{state} information. Directional cues use spatial correspondence (e.g., left motor for left error) and co--activation for diagonals \cite{Oyarzun2019VibrotactilePuzzle}. To reduce cognitive load, we adopt pull--oriented, pulsed bursts (200~ms on / 200~ms off) rather than sustained vibration, reflecting evidence that pull metaphors lower mental demand and improve completion time \cite{Smith2023PullPush}. To limit adaptation, directional pulses run at half intensity, while \emph{state} cues (within tolerance) trigger a brief, full--intensity all--motors burst. Placement of the controller housing on the dorsal wrist mirrors effective dorsal surface usage for skin--stretch \cite{Li2025SkinStretch} and avoids interference with sleeves or gowns.

\subsection{XR Software Pipeline}
We implemented the AR application in Unity (MRTK). A virtual 10~mm target sphere and a depth reference are rendered at set depths (30--40~cm). The stylus pose is tracked at 120~Hz (OptiTrack) and streamed to Unity. Axis--wise error between the tool tip and target triggers directional cues whenever magnitude exceeds \(\pm2\)~mm. If error remains within a spherical tolerance for 500~ms, a success state cue fires. For visual baselines, we optionally render the real stylus mesh and use opaque targets to mirror high--veridical conditions reported by Yang et~al.\ \cite{Yang2020DepthPerception}. The software is modular to accommodate future skin--stretch or audio--haptic channels \cite{Lu2024SonoHaptics}.

\section{User Studies}
\subsection{Participants}
We ran two pre--registered studies approved by the Fictional University IRB. \textbf{Study~1} enrolled 21 participants (11 female, 10 male; ages 20--35) with normal or corrected vision and no prior exposure to our device. \textbf{Study~2} enrolled 27 participants (16 female, 11 male; ages 22--37), including 10 surgical residents and 17 graduate students; five participants took part in both studies. All provided consent and received compensation.

\subsection{Study 1: Cue Identification Under Cognitive Load}
Participants wore HoloLens~2 and the wristband while completing a dual--task: a letter--count visual search (2~s) followed by a 2~s haptic window. They verbally identified one of five cue types per trial: four directions (left/right/up/down) and a success state (all motors). Each cue appeared 10 times (50 trials total) in randomized order. The protocol follows dual--task paradigms known to reveal vibrotactile discriminability limits and workload interactions \cite{Weber2008VibrotactileGuidance}. We measured accuracy and response time; we also logged confusion matrices to characterize specific misclassifications (e.g., up vs.\ down).

\subsection{Study 2: Multimodal Tool Guidance}
We compared three within--subjects conditions with counterbalanced order: \emph{AR--only} (opaque target and rendered real stylus, no haptics), \emph{Haptic--only} (no virtual renderings; guidance via wrist cues and a physical depth reference), and \emph{Multimodal} (AR + wrist haptics). Targets varied in depth (30, 35, 40~cm) and lateral offset (\(\pm10\)~mm). Each participant completed 18 trials per condition (6 positions \(\times\) 3 repetitions) after a brief training block. Dependent measures were Euclidean alignment error, completion time, and overshoot frequency. After each condition, participants completed NASA--TLX and SUS. We also collected post--study rankings and interviews. The AR--only baseline mirrors the most effective visual settings in \cite{Yang2020DepthPerception}; the Haptic--only and Multimodal conditions probe whether peripheral cues can match or exceed visual performance and how the two channels integrate in practice \cite{Wang2025Teleoperation,Smith2023PullPush,Lu2024SonoHaptics}.

\subsection{Hypotheses and Metrics}
We pre--registered: \textbf{H1} (Perceptual accuracy): pull--style wrist cues will yield $>\!90\%$ identification accuracy under load, with most confusions occurring along the vertical axis, as documented for proximal tactors \cite{Weber2008VibrotactileGuidance,Smith2023PullPush}. \textbf{H2} (Multimodal precision): Multimodal feedback will reduce depth error relative to AR--only by mitigating overshoot and lateral drift \cite{Yang2020DepthPerception}. \textbf{H3} (Speed--accuracy trade--off): Wrist haptics will reduce error but may increase completion time versus AR--only, reflecting observed trade--offs in teleoperation \cite{Wang2025Teleoperation}. \textbf{H4} (Workload and usability): Multimodal will produce lower NASA--TLX (mental demand) and higher SUS than single--modality conditions, consistent with offloading visual demand \cite{Oyarzun2019VibrotactilePuzzle,Pezent2021Tasbi}.

\section{Findings and Analysis}
\subsection{Haptic Cue Comprehension Under Load}

\paragraph{Accuracy, confusions, and reliability.}
Participants identified cues with high reliability (overall $92\%$). Directional cues were recognized $89\%$ ($\mathrm{SD}=7\%$), whereas the state cue reached $98\%$ ($\mathrm{SD}=3\%$). A repeated--measures ANOVA showed a significant effect of cue type on accuracy ($F(4,80)=5.13$, $p<0.01$). Post--hoc tests revealed that \emph{up} ($82\%$) and \emph{down} ($84\%$) were less accurate than \emph{left} ($93\%$) and \emph{right} ($94\%$), indicating a vertical--axis ambiguity consistent with earlier reports on placement and skin mechanics \cite{Weber2008VibrotactileGuidance}. Mean response time was $1.1\,\mathrm{s}$ ($\mathrm{SD}=0.3\,\mathrm{s}$) with no reliable differences across cue classes. The overall pattern supports the use of simple spatial mappings for direction and a single high--salience pattern for confirmation.

\begin{figure}[t]
  \centering
  \begin{tikzpicture}
    \begin{axis}[
      ybar,
      ymin=0, ymax=100,
      bar width=12pt,
      width=0.9\linewidth, height=6cm,
      ylabel={Accuracy (\%)},
      symbolic x coords={Left,Right,Up,Down,State},
      xtick=data,
      ymajorgrids=true,
      grid style=dashed,
    ]
      \addplot+[error bars/.cd, y dir=both, y explicit]
        coordinates {
          (Left,93) +- (0,7)
          (Right,94) +- (0,7)
          (Up,82) +- (0,7)
          (Down,84) +- (0,7)
          (State,98) +- (0,3)
        };
    \end{axis}
  \end{tikzpicture}
  \caption{Cue identification accuracy by type (mean $\pm$ SD). Vertical cues underperform horizontal ones; the state cue is near ceiling \cite{Weber2008VibrotactileGuidance,Smith2023PullPush}.}
  \label{fig:cue-accuracy}
\end{figure}
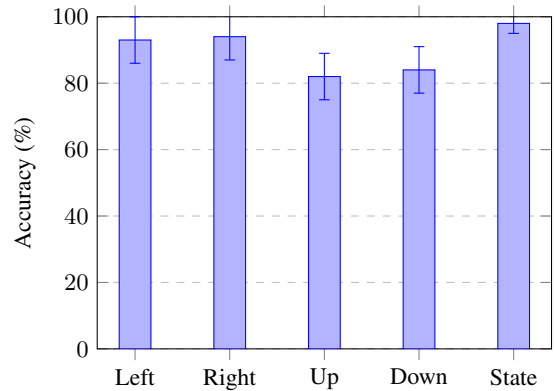

\paragraph{Learning, robustness, and ceiling effects.}
To probe learning effects, we examined trial order; the vertical $>$ horizontal confusion asymmetry persisted across the session, suggesting a spatial--encoding source rather than a first--exposure artifact. The state cue’s near--ceiling accuracy across the entire session indicates that a brief, full--band activation is a robust pattern for \emph{confirmation} even under dual--task load \cite{Weber2008VibrotactileGuidance,Smith2023PullPush}. In practical terms, this supports the use of a single, unmistakable success signal and discourages overloading the codebook with multiple “success” variants.

\begin{figure}[t]
  \centering
  \begin{tikzpicture}
    \begin{axis}[
      ybar,
      ymin=0,
      bar width=12pt,
      width=0.9\linewidth, height=5.5cm,
      ylabel={Mean RT (s)},
      symbolic x coords={Left,Right,Up,Down,State},
      xtick=data,
      ymajorgrids=true,
      grid style=dashed,
    ]
      \addplot coordinates {(Left,1.1) (Right,1.1) (Up,1.1) (Down,1.1) (State,1.1)};
    \end{axis}
  \end{tikzpicture}
  \caption{Response time by cue type (means). No reliable differences across cue classes; participants respond uniformly after learning the mapping.}
  \label{fig:cue-rt}
\end{figure}
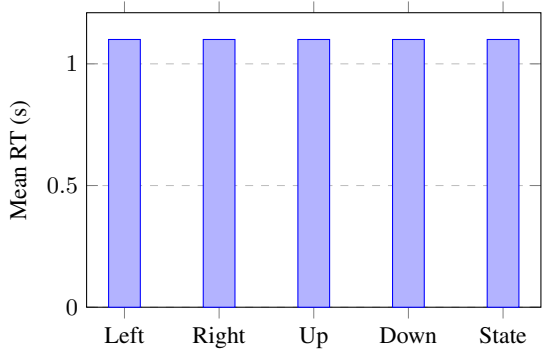

\paragraph{Design takeaways for cue coding.}
Two implications follow. First, vertical discrimination benefits from (i) slight spatial relocation (e.g., increased dorsal/volar separation) or (ii) a mixed code (spatial $+$ short temporal ornament) to break symmetry. Second, confirmation cues should remain brief and intense to prevent adaptation while providing unmistakable closure. These choices align with “pull” metaphors and short pulses that reduce cognitive load \cite{Smith2023PullPush}.

\subsection{Guided Tool Alignment}

\begin{figure}[t]
  \centering
  \begin{tikzpicture}
    \begin{axis}[
      ybar,
      ymin=0,
      bar width=16pt,
      width=0.9\linewidth, height=6cm,
      ylabel={Mean Error (mm)},
      symbolic x coords={AR-only,Haptic-only,Multimodal},
      xtick=data,
      ymajorgrids=true,
      grid style=dashed,
    ]
      \addplot+[error bars/.cd, y dir=both, y explicit]
        coordinates {
          (AR-only,8.4) +- (0,2.1)
          (Haptic-only,7.5) +- (0,2.0)
          (Multimodal,5.8) +- (0,1.6)
        };
    \end{axis}
  \end{tikzpicture}
  \caption{Alignment error by condition (mean $\pm$ SD). \emph{Multimodal} significantly reduces depth error vs.\ single--modality baselines \cite{Yang2020DepthPerception}.}
  \label{fig:error}
\end{figure}
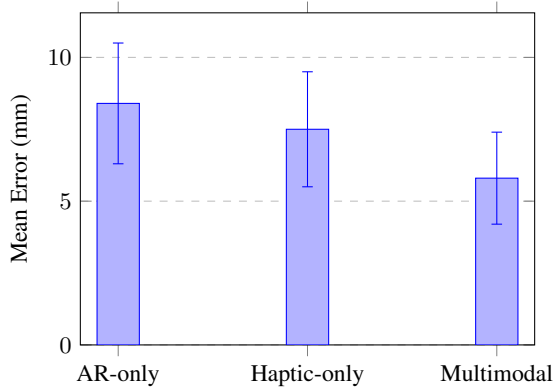

\paragraph{Depth error and error profiles.}
Figure~\ref{fig:error} shows a strong effect of condition on depth error ($F(2,52)=16.8$, $p<0.001$): \emph{Multimodal} (Mean $=5.8$\,mm, $\mathrm{SD}=1.6$) outperformed both \emph{AR--only} (Mean $=8.4$\,mm, $\mathrm{SD}=2.1$, $p<0.01$) and \emph{Haptic--only} (Mean $=7.5$\,mm, $\mathrm{SD}=2.0$, $p<0.05$); the latter two did not differ significantly. The pattern mirrors prior OST--AR findings: when visual occlusion or registration is imperfect, auxiliary haptics damp overshoot and drift \cite{Yang2020DepthPerception}.

\paragraph{Completion time and the speed--accuracy frontier.}
\emph{Multimodal} trials were slightly slower (Mean $=9.2$\,s, $\mathrm{SD}=1.8$) than \emph{AR--only} (Mean $=8.0$\,s, $\mathrm{SD}=1.5$) and \emph{Haptic--only} (Mean $=7.8$\,s, $\mathrm{SD}=1.4$); the \emph{Multimodal} vs.\ \emph{AR--only} difference was significant ($p<0.05$). Participants reported deliberate cross--checks of visual and haptic information. This mirrors teleoperation findings where wrist--level feedback improves precision yet nudges movement time upward, a pragmatic speed--accuracy trade--off that can be tuned by users or policies \cite{Wang2025Teleoperation}.

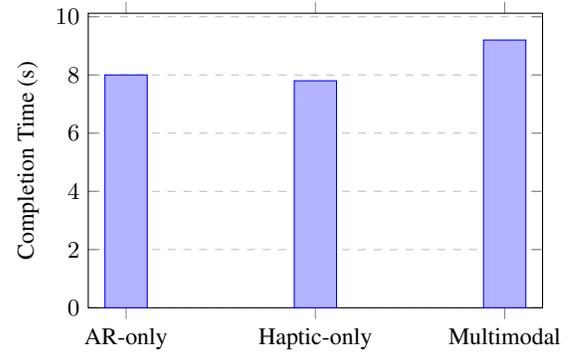
\begin{figure}[t]
  \centering
  \begin{tikzpicture}
    \begin{axis}[
      ybar,
      ymin=0,
      bar width=16pt,
      width=0.9\linewidth, height=5.5cm,
      ylabel={Completion Time (s)},
      symbolic x coords={AR-only,Haptic-only,Multimodal},
      xtick=data,
      ymajorgrids=true,
      grid style=dashed,
    ]
      \addplot coordinates {(AR-only,8.0) (Haptic-only,7.8) (Multimodal,9.2)};
    \end{axis}
  \end{tikzpicture}
  \caption{Completion time by condition. \emph{Multimodal} introduces a small time premium consistent with increased confirmation checks \cite{Wang2025Teleoperation}.}
  \label{fig:time}
\end{figure}

\paragraph{Overshoot control and stability.}
Overshoot frequency---entering beyond target depth---was highest in \emph{Haptic--only} ($27\%$) and lowest in \emph{Multimodal} ($9\%$). This drop is consistent with the complementary value of concurrent channels in ambiguous depth contexts \cite{Yang2020DepthPerception}. (AR--only overshoot was not reported and is therefore omitted from figures and inferential comparisons.)

\subsection{Subjective Experience}

\paragraph{Usability and workload.}
SUS differed across conditions ($F(2,52)=20.3$, $p<0.001$): \emph{Multimodal} scored highest ($88.1/100$), followed by \emph{AR--only} ($78.7$) and \emph{Haptic--only} ($70.4$). NASA--TLX indicated the lowest workload for \emph{Multimodal} (Mean $=35/100$) and the highest for \emph{Haptic--only} ($58$). Participants frequently described the pulses as a gentle \emph{pull} toward the target, echoing evidence that pull metaphors reduce mental demand and aid control \cite{Smith2023PullPush}. Preferences for a lightweight wrist form factor and embodied spatial mapping align with prior AR wristband results \cite{Oyarzun2019VibrotactilePuzzle}.

\begin{figure}[t]
  \centering
  \begin{tikzpicture}
    \begin{axis}[
      ybar,
      ymin=0, ymax=100,
      bar width=16pt,
      width=0.9\linewidth, height=5.5cm,
      ylabel={SUS (0--100)},
      symbolic x coords={AR-only,Haptic-only,Multimodal},
      xtick=data,
      ymajorgrids=true,
      grid style=dashed,
    ]
      \addplot coordinates {(AR-only,78.7) (Haptic-only,70.4) (Multimodal,88.1)};
    \end{axis}
  \end{tikzpicture}
  \caption{System Usability Scale (SUS) by condition. \emph{Multimodal} is rated most usable.}
  \label{fig:sus}
\end{figure}
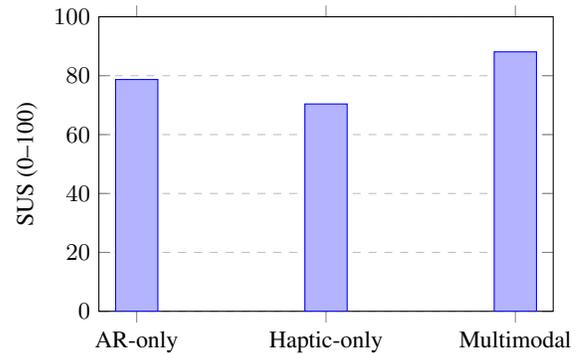

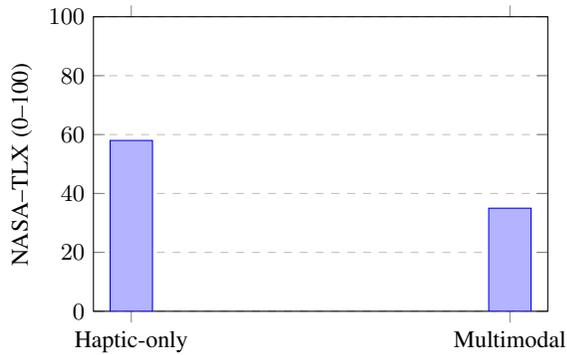
\begin{figure}[t]
  \centering
  \begin{tikzpicture}
    \begin{axis}[
      ybar,
      ymin=0, ymax=100,
      bar width=16pt,
      width=0.9\linewidth, height=5.5cm,
      ylabel={NASA--TLX (0--100)},
      symbolic x coords={Haptic-only,Multimodal},
      xtick=data,
      ymajorgrids=true,
      grid style=dashed,
    ]
      \addplot coordinates {(Haptic-only,58) (Multimodal,35)};
    \end{axis}
  \end{tikzpicture}
  \caption{NASA--TLX by condition (only reported values shown). \emph{Multimodal} yields the lowest workload.}
  \label{fig:tlx}
\end{figure}

\paragraph{Qualitative synthesis.}
Participants valued the wristband’s unobtrusiveness (no finger encumbrance) and the intuitive “tapped/pulled” feeling for direction. The main complaint concerned \emph{up} vs.\ \emph{down} confusion---consistent with the quantitative gap and prior work \cite{Weber2008VibrotactileGuidance}. Clinicians highlighted use cases where overlays may be occluded (e.g., blood/tissue) or rendering sterile tools is restricted; in such cases, peripheral haptics provide redundancy that preserves confidence \cite{Yang2020DepthPerception}. Several asked whether skin--stretch or squeeze could further disambiguate vertical cues; dorsal skin--stretch can improve spatial understanding relative to vibration \cite{Li2025SkinStretch}, and squeeze can increase perceived realism \cite{Pezent2021Tasbi}. These modalities are promising \emph{additions} rather than replacements.

\subsection{Design--Centric Reading and Practical Guidance}

\paragraph{Coding policy.}
Use spatial correspondence for direction; encode confirmations with brief, high--intensity all--motor bursts to ensure unmistakable closure and minimize adaptation. Favor pull--style, pulsed patterns for direction to reduce mental demand and speed decisions \cite{Smith2023PullPush}.

\paragraph{When to go multimodal.}
Adopt multimodality when visual reliability is imperfect (glare, occlusion, tracking loss). The reduction in overshoot and error justifies a small time premium. Provide user controls (gain, duty cycle, auto--fade) so experts can trade speed for assurance \cite{Wang2025Teleoperation}.

\paragraph{Path to greater precision.}
For vertical disambiguation, consider dorsal or forearm placements and mixed spatial--temporal patterns; explore ring actuators for local precision \cite{Nguyen2025PinchRing} and add light squeeze or skin--stretch to encode graded certainty \cite{Pezent2021Tasbi,Li2025SkinStretch}. Audio--haptic mappings (e.g., size$\rightarrow$pitch, certainty$\rightarrow$amplitude) may further reduce visual reliance in cluttered scenes \cite{Lu2024SonoHaptics}.

\bibliographystyle{abbrv-doi-hyperref-narrow}

\bibliography{template}
\clearpage
\onecolumn
\clearpage

\end{document}